\documentclass[aps,pra,reprint,superscriptaddress,flushbottom]{revtex4-1}

\usepackage{amsmath}
\usepackage{amssymb}
\usepackage{amsthm}

\usepackage{setspace}
\long\def\symbolfootnote[#1]#2{\begingroup\def\thefootnote{\fnsymbol{footnote}}\footnote[#1]{#2}\endgroup}
\usepackage{amscd}

\usepackage{color}

\usepackage{float}

\newcommand{\bra}[1]{\langle {#1} \vert}
\newcommand{\ket}[1]{\vert {#1} \rangle}

\newcommand{\Svn}{S_{\text{vN}}}
\newcommand{\Sacc}{S_{\text{acc}}}
\newcommand{\Sfluc}{S_{\text{fluct}}}

\usepackage{epsfig}

\begin{document}
\title{Generating accessible entanglement in bosons via pair-correlated tunneling}
\author{T.\,J.~Volkoff}
\email{volkoff@konkuk.ac.kr}
\affiliation{Department of Physics, Konkuk University, Seoul 05029, Korea}
\author{C.\,M.~Herdman}
\email{cherdman@middlebury.edu}
\affiliation{Department of Physics, Middlebury College, Middlebury, VT, 05753, USA}

\begin{abstract}
We consider an extended Bose-Hubbard model that includes pair-correlated tunneling. We demonstrate that a minimal four-mode implementation of this model exhibits a pair-correlated regime in addition to Mott insulator and superfluid regimes. We propose a low complexity variational subspace for the ground state of the system in the pair-correlated regime, which we find to be numerically exact in pure pair-tunneling limit. Additionally, we propose a parameter-free high fidelity model wave function that qualitatively captures the features of the ground state in the pair-correlated regime. Although the operationally accessible entanglement vanishes deep inside the Mott insulator and superfluid regimes due to particle number conservation, we find that in the pure pair-correlated tunneling limit the accessible entanglement entropy grows logarithmically with the number of particles. Furthermore, we demonstrate that upon application of a unitary beamsplitter operation, the pair-correlated ground state is transformed into a state with completely accessible entanglement that is not limited by super-selection rules.
\end{abstract}
\maketitle

\section{Introduction}
A variety of experimental systems of trapped strongly interacting bosons are accurately described by Bose-Hubbard (BH) model of itinerant bosons hopping between localized modes. Such quantum phases of strongly correlated systems of bosons have been experimentally shown to display such quantum properties as stable superfluid flow \cite{minguzzi,cohensfflow,PhysRevLett.110.025301,PhysRevLett.106.130401}, quantized circulation \cite{PhysRevLett.99.260401,pitaevskiibook}, and spin-squeezing \cite{oberthaler,PhysRevLett.113.103004,ciraczoller,smerzicounter}. Additionally, strongly correlated bosonic ground states arising from Bose-Hubbard dynamics possess quantum entanglement that can be harvested for quantum circuit-based quantum information processing protocols \cite{greiner,melkoherdman} and quantum metrology \cite{treutlein,nongausscoldatommetro}.

The entanglement between two subsystems of a pure quantum state $| \psi \rangle$ may be quantified by the bipartite von Neummann entanglement entropy
$$\Svn\bigl(\ket{\psi}\bigr)=-\text{Tr}\rho_A \ln \rho_A$$
where $\rho_A$ is the reduced density matrix under a bipartition into subsystems $A$ and its complement $B$: $\rho_A = \mathrm{Tr}_B | \psi \rangle \langle \psi |$. However, superselection rules constrain the entanglement that is operationally accessible via local operations in the presence of particle number conservation in systems of non-relativistic bosons \cite{wiseman,RevModPhys.79.555}. Wiseman and Vaccaro defined the accessible entanglement entropy:
\begin{equation}
\Sacc\bigl( \ket{\psi} \bigr) = \sum_{n =0}^{N}p_{n}\Svn \bigl( P_{n} \ket{\psi}/\sqrt{p_n} \bigr)
\label{eqn:opent}
\end{equation}
where $P_{n}$ is the projection onto the subspace where $n$ particles are in the $A$ subsystem, and $p_n = \bra{\psi} P_n \ket{\psi}$. Accordingly, the total entanglement entropy may be decomposed into contributions from particle number fluctuations between subsystems, $\Sfluc( \ket{\psi} ) = - \sum_{n =0}^{N} p_n \ln p_n$, and the accessible entanglement: $\Svn = \Sfluc + \Sacc$ \cite{klichlevitov}.

In the minimal BH system with two modes, under a mode bipartition there is only a single mode in each subsystem, and thus all entanglement is due to fluctuations of particles between subsystems; correspondingly, there is no entanglement that is accessible via local operations. In fact, at least four single particle modes are required to allow for non-zero accessible entanglement necessary for meaningful entanglement distribution and concentration\cite{wiseman,wisemanexamples}. However, in both the non-interacting and strongly interacting limits of the BH model with at least four single particle modes, the accessible entanglement vanishes; specifically, the ground state in the Mott insulator regime limits to an unentangled product state, whereas in the non-interacting limit, the ground state is a Bose-Einstein condensate where all entanglement is due to fluctuations.

In this paper we demonstrate how pair-correlated dynamics can generate accessible entanglement in itinerant boson systems described by an extended BH model. We consider a minimal four mode model of $N$ spinless bosons defined by the Hamiltonian $H=H_{\text{BH}}+H_{\text{pair}}$, where $H_{\text{BH}}$ is the Bose-Hubbard Hamiltonian for four single-particle modes that represent, e.g., the sites of an optical ring lattice, and $H_{\text{pair}}$ describes pair-correlated hopping dynamics of the particles. Explicitly,
\begin{align}
H=\sum_{j=0}^{3}\biggl[ \biggr.{U\over 2}n_{j}\left( n_{j} - 1 \right)  &- J\left( a_{j+1}^{\dagger}a_{j}  +h.c.\right)  \nonumber \\ &- \bigl. T_{2}\left( a_{j+1}^{\dagger 2}a_{j}^{2} + h.c. \right) \biggr]
\label{eqn:ham}
\end{align}
where $a_j$ is the bosonic annihilation operator in mode $j$, $n_j = a_j^\dagger a_j$, and the first two terms corresponds to $H_{\text{BH}}$ and the third term to $H_{\text{pair}}$. All parameters $U$, $T_{2}$, $J$ are taken to be non-negative.

Coherent pair tunneling in the presence of single-particle tunneling suppression has been observed in strongly-coupled, optical double-well systems of ultracold $^{87}$Rb \cite{blochfolling}. For spinless bosons, pair tunneling is analogous to the superexchange phenomenon in magnetic systems \cite{auerbach,superex}. In ultracold spinor gases of $^{87}$Rb, both pair tunneling processes and photon-assisted hyperfine superexchange can be controlled by modulation of an optical superlattice \cite{PhysRevLett.107.210405,PhysRevA.77.063603,PhysRevA.80.013605}. Singlet pair tunneling in spinor Bose gases can be treated in the same way as the analysis of pair tunneling in the present work because, e.g., $a^{2}$ and $a_{\uparrow,i}a_{\downarrow,j}$ are both $\mathfrak{su}(1,1)$ ladder operators. Quantum coherence due to pair-correlated tunneling in many-body, two-mode bosonic systems has previously been shown to be useful for near-optimal quantum estimation of single particle tunneling amplitudes \cite{volkoffmetro}. In this paper, we address how pair-correlated tunneling can generate accessible entanglement and, when combined with an implementation of a matter-wave beamsplitter, allows the conversion of fluctuation entropy into accessible entanglement.

The regime of $H$ defined by $T_{2}\gg U, J$, which we call the PC (pair-correlated) regime, corresponds to an ``untwisting'' of the twisted superfluid phase, the latter defined by $U$, $J>0$, $T_{2}<0$, and so named due to the alternating sign of the argument of the correlation function $\langle a_{j+1}^{\dagger} a_{j} \rangle$  \cite{luhmann,jurgensen}. For $J=0$, the untwisting operation that changes the sign of $T_{2}$ is implemented by an on-site, alternating phase shift given in Eq.(\ref{eqn:onsiteunitary}). However, for $J\neq 0$, there is no local U(1) rotation, i.e., generated by $\sum_{j=0}^{3}\theta_{j}n_{j}$, that changes the sign of $T_{2}$ while keeping the sign of $J$ fixed. Therefore, the twisted superfluid and PC regimes are not generically connected by local operations.

To analyze the PC regime, we first identify a low-complexity subspace where the ground state in the pure pair-tunneling ($J=U=0$) limit resides. Additionally, we introduce a parameter-free high-fidelity model wave function that quantitatively describes the ground state in the PC regime. Unlike deep in the Mott insulator and superfluid regimes, in the pure pair-tunneling limit deep inside the PC regime, we show that the ground state has non-vanishing  accessible entanglement that scales as $\Sacc \sim \ln N$. By exploiting an experimentally-realizable nonlocal unitary operation, viz., a 50:50 matter wave beamsplitter that hybridizes the single particle modes, we demonstrate an ``entanglement switch" that increases the coefficient of the logarithmic scaling of accessible entanglement in the ground state of the PC regime, resulting in a many-boson state with completely accessible entanglement. By demonstrating that the PC regime exhibits useful and manipulable entanglement, we establish few-mode coherent pair hopping as an elementary module for bosonic quantum information processing.

Deep in the PC regime, the ground state is well approximated by the ground state of the interaction $H_{\text{pair}}$. Unlike the Mott insulating regime ($U\gg T_{2}$, $U\gg J$) or the superfluid regime ($J\gg T_{2}$, $J\gg U$), for which the ground states are easily obtained from perturbation theory (in fact, for $T_{2}=0$ the system is solvable by algebraic Bethe ansatz \cite{linksfourmode,foerster1,enolskii1,enolskii2}) the PC regime does not admit a clear method to obtain an analytical ground state. In fact, $H_{\text{pair}}$ can be written as
\begin{equation}
H_{\text{pair}} = H_{R}-a_{0}^{\dagger 2}a_{2}^{2}-a_{2}^{\dagger 2}a_{0}^{2}-a_{1}^{\dagger 2}a_{3}^{2}-a_{3}^{\dagger 2}a_{1}^{2}, \nonumber
\end{equation}
where $H_{R}=\sum_{j,j'}a_{j}^{\dagger 2}a_{j'}^{2}$ is an exactly solvable Richardson model \cite{richardson}. Therefore, $H_{\text{pair}}$ can be considered as a large, nonlinear perturbation of an exactly solvable model, although the two-axis countertwisting Hamiltonian, which is the two-mode analog of $H_{\text{pair}}$, is exactly solvable \cite{draayer1,draayer2}.

\section{\label{sec:pcssec}Pair-correlated ground state} To gain insight into the structure of the ground state of $H_{\text{pair}}$, we first note that $W^{\dagger}H_{\text{pair}}W = -H_{\text{pair}}$, where $W$ is the unitary transformation
\begin{equation}
\label{eqn:onsiteunitary}
W=\exp \left[ -i {\pi \over 4}\left( n_0 - n_1 + n_2 - n_3 \right) \right].
\end{equation}
Note that for any quantum state $\rho$ of $N$ bosons in four modes, $W\rho W^{\dagger} = V\rho V^{\dagger}$, where $V$ can be taken to be $\exp{[i{\pi \over 2}(n_1 + n_3) ] }$ or $\exp{[ -i{\pi \over 2} (n_0 + n_2 ) ]}$.  Due to this discrete antisymmetry of $ H_{\text{pair}}$, we expect the ground state to obey $W^{2}\ket{\Psi_{0}} = \ket{\Psi_{0}}$. Furthermore, $H_{\text{pair}}$ is invariant under the dihedral group $D_{8}$ generated by cyclic permutation of the modes $a_{0}\rightarrow a_{1} \rightarrow a_{2} \rightarrow a_{3}\rightarrow a_{0}$ and the transposition $a_{0}\leftrightarrow a_{2}$. Notice that, by considering the action of these symmetry operations on the vector of pair annihilation operators $(a_{0}^{2},\ldots, a_{3}^{2})$, the full symmetry group is found to be given by the semidirect product $D_{8} \rtimes \mathbb{Z}_{2}$. In postulating variational ansatze for the ground state of $H_{\text{pair}}$, we are motivated by the fact that quantum states that exhibit pair correlations have been used to analyze interacting bosonic systems since the early days of the quantum theory of superfluidity \cite{valatin,girardeau}, and have recently been utilized to rigorously formulate a number-conserving version of the Bogoliubov theory \cite{caveszhang,caveszhang2}.

\begin{figure}
\includegraphics[width= \columnwidth]{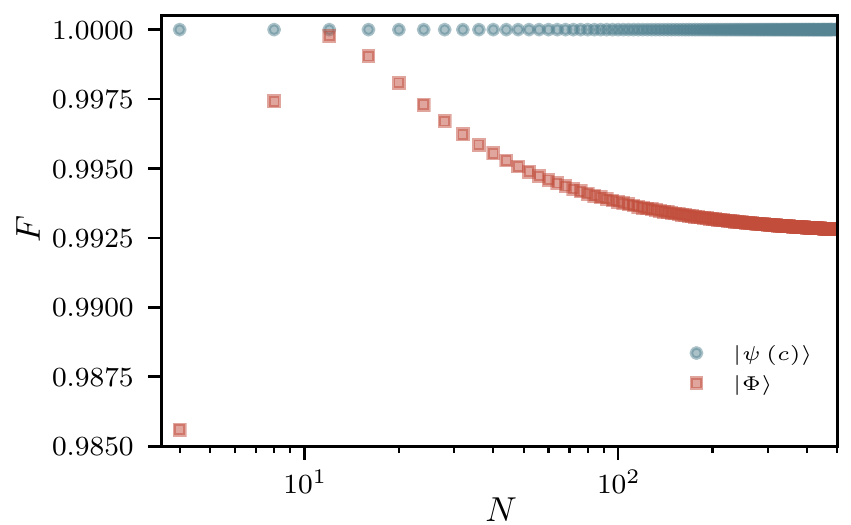}\llap{\makebox[6.5cm][l]{\raisebox{0.85cm}{\includegraphics[height=2.4cm]{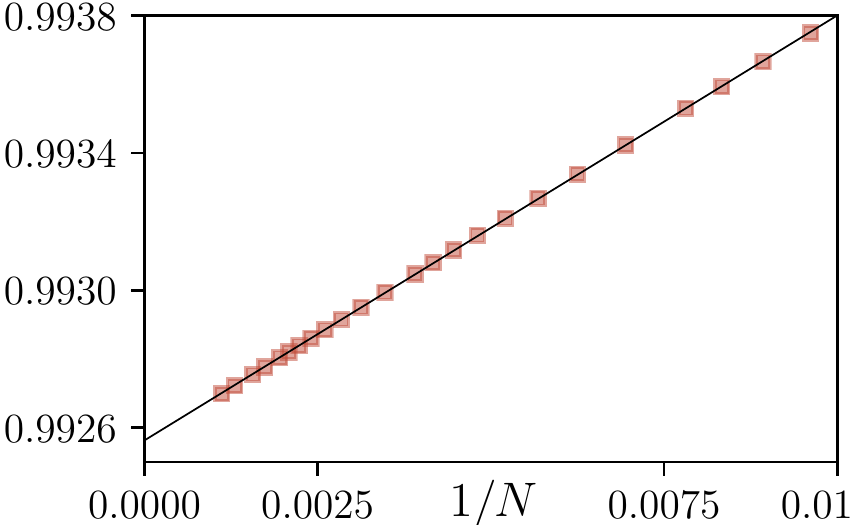}}}}
\caption{\label{fig:fig1} Optimized fidelity $F$ of the variational state $\ket{ {\psi(c)}}$ given in Eqn. \ref{eqn:optstate}, and fidelity of the model ground state $\ket{\Phi} $ given in Eqn. \ref{eqn:modelvar} to the exact ground state for pure pair-tunneling $\ket{\Psi_0}$, as computed by numerical diagonalization of Eqn. \ref{eqn:ham}.  with $U=J=0$. {\it Inset}: Fidelity of $\ket{\Phi} $ vs. $1/N$. The line is a linear (in $1/N$) fit, that provides an estimate of the $N=\infty$ fidelity $F_\infty \simeq 0.99256$.}
\end{figure}

We expect that the $N$-particle ground state of $H_{\text{pair}}$ is in a variational subspace that is invariant under all symmetry operations of $D_{8} \rtimes \mathbb{Z}_{2}$. By defining $M=N/2$ and $k_{\ell}=2\pi \ell / M$, we propose the following ansatz, $\ket{\psi(c)}$  as a variational ground state of $H_{\text{pair}}$:
\begin{align}
\ket{\psi (c) }&= \sum_{\ell=0}^{\left \lfloor{ M\over 2}\right \rfloor }c_{\ell} \ket{\varphi_{\ell}}, \nonumber \\
\ket{\varphi_{\ell}} &={1\over \mathcal{N}_{\ell}}\left[ \vphantom{a\over b}\left( a_{0}^{\dagger 2}+e^{ik_{\ell}}a_{1}^{\dagger 2}+a_{2}^{\dagger 2}+e^{ik_{\ell}}a_{3}^{\dagger 2} \right)^{M} \right.\label{eqn:optstate} \\
&+ \left. \left( a_{0}^{\dagger 2}+e^{-ik_{\ell}}a_{1}^{\dagger 2}+a_{2}^{\dagger 2}+e^{-ik_{\ell}}a_{3}^{\dagger 2} \right)^{M}  \vphantom{a\over b} \right] \ket{0,0,0,0} \nonumber
\end{align}
where $c_\ell$ are real variational parameters, and $ \ket{\varphi_{\ell}}$ are normalized, but non-orthogonal states (refer to Appendix \ref{sec:app1} for calculation of the normalization constants $\mathcal{N}_{\ell}$). For $N=2$, $\ket{\varphi_{0}}$ is the exact ground state of $H_{\text{pair}}$ corresponding to energy eigenvalue $E_{0}=-4$. Similarly, for $N=4$, the exact ground state (up to normalization) corresponds to $c_{0}=1$ and $c_{1}=-3+2\sqrt{2}$ with energy eigenvalue $E_{0}=-8\sqrt{2}$.
Although we have not proven that the ground state of $H_{\text{pair}}$ takes the form of Eq.(\ref{eqn:optstate}) for all $N$, we do not find any physically meaningful deviation of the optimal analytical state Eq.(\ref{eqn:optstate}) from the numerically calculated ground state; Fig. \ref{fig:fig1} shows that the optimal variational state $\ket{\psi(c)}$ has numerically perfect (up to machine precision) optimal fidelity $F = \max_{c}| \langle \psi(c) | \Psi_0 \rangle|$  to the pure pair-tunneling ($U=J=0$) ground state $\ket{\Psi_0}$ up to $N=\mathcal{O}(10^2)$.

To provide additional insight into the PC regime, we analyze a unitary transformation $\mathcal{V}H_{\text{pair}}\mathcal{V}^{\dagger}= \tilde{H}_{\text{pair}} $ that allows one to construct an approximate linear quantum dynamics for this regime. For even $N$, the construction involves identifying the ground state subspace of $\tilde{H}_{\text{pair}} $ with a spin-$N/4$ representation of $SU(2)$. To see this, consider the unitary operator:
\begin{equation}
\mathcal{V}=e^{{i\pi\over 4}(a_{0}^{\dagger}a_{2}+h.c.)}e^{{i\pi\over 4}(a_{1}^{\dagger}a_{3}+h.c.)} \label{eqn:Vop}
\end{equation}
which performs a mode transformation into the modes $c_{\tilde{j}}^{\dagger} = \mathcal{V}a_{j}^{\dagger}\mathcal{V}^{\dagger}=({a_{j}^{\dagger}+ia_{j+ 2}^{\dagger}) / \sqrt{2}}$, where $j+2$ is understood modulo 4. In terms of the $c_{\tilde{j}}$ operators, $\tilde{H}_{\text{pair}}$ takes the form
\begin{equation}\tilde{H}_{\text{pair}}= \mathcal{V}H_{\text{pair}}\mathcal{V}^{\dagger} = -4T_{2}\left( c_{\tilde{0}}^{\dagger}c_{\tilde{2}}^{\dagger}c_{\tilde{1}}c_{\tilde{3}} + h.c. \right).
\label{eqn:ch}
\end{equation}
Note that with
\begin{equation}
T^{(0,1)}=e^{i{\pi\over 2}\left(n_{\tilde{0}}-n_{\tilde{1}}\right)}, \nonumber
\end{equation}
it follows that $T^{(0,1)}\tilde{H}_{\text{pair}}T^{(0,1)\dagger}=-\tilde{H}_{\text{pair}}$, which implies that a local rotation (with respect to the mode bipartition $\lbrace \tilde{0},\tilde{1} \rbrace \sqcup \lbrace \tilde{2},\tilde{3}  \rbrace $) in the $c_{\tilde{j}}$ basis  can change the sign of $\tilde{H}_{\text{pair}}$. By expressing $\tilde{H}_{\text{pair}}$ as in Eq. (\ref{eqn:ch}), it is clear that the algebra generated by the observables $(n_{\tilde{0}} + n_{\tilde{1}})$
and $(n_{\tilde{2}} + n_{\tilde{3}})$ (or, equivalently, by $(n_{\tilde{0}} + n_{\tilde{1}})$ and the identity) consists of conserved quantities. Because of the permutation symmetry of $\tilde{H}_{\text{pair}}$, the ground state of $\tilde{H}_{\text{pair}}$ lies in the $(M+1)^{2}$-dimensional subspace spanned by
\begin{equation}
\biggl\lbrace \ket{ r,M-r,s,M-s } \biggr\rbrace_{r,s\in \lbrace 0,\ldots , M\rbrace}.\nonumber
\end{equation}
However, the algebra generated by $(n_{\tilde{0}}-n_{\tilde{2}})$ and $(n_{\tilde{1}}-n_{\tilde{3}})$ also consists of observables that commute with $\tilde{H}_{\text{pair}}$. Therefore, the ground state lies in the $(M+1)$-dimensional subspace
\begin{equation}
V= \biggl\lbrace \ket{ s,M-s,s,M-s } \biggr\rbrace_{ r,s\in \lbrace 0,\ldots , M\rbrace}.\nonumber
\end{equation}

From the action of $c_{\tilde{j}}$ and $c_{\tilde{j}}^{\dagger}$ on the symmetric Fock space, it follows that $c_{\tilde{0}}^{\dagger}c_{\tilde{2}}^{\dagger}c_{\tilde{1}}c_{\tilde{3}} \big\vert_{V}$ is equal, as a linear operator, to the spin observable
\begin{equation}
\sqrt{{M\over 2}+J_{z}}J_{+}\sqrt{{M\over 2} - J_{z}} \nonumber
\end{equation}
acting on a spin-$M/2$ representation of $\mathfrak{su}(2)$ and that $c_{\tilde{1}}^{\dagger}c_{\tilde{3}}^{\dagger}c_{\tilde{0}}c_{\tilde{2}} \big\vert_{V}$ is equal, as a linear operator, to the spin observable
\begin{equation}
\sqrt{{M\over 2}-J_{z}}J_{-}\sqrt{{M\over 2} + J_{z}}\nonumber
\end{equation}
acting on the spin-$M/2$ representation. Therefore, $\tilde{H}_{\text{pair}}$ is proportional, as a linear operator, to
\begin{equation}
\sqrt{{M\over 2}+J_{z}}J_{+}\sqrt{{M\over 2} - J_{z}} + \sqrt{{M\over 2}-J_{z}}J_{-}\sqrt{{M\over 2} + J_{z}}
\label{eqn:Hpairspin}
\end{equation}
acting in a spin-$N/4$ representation of $SU(2)$. As a result, one finds that the Hamiltonian $-\tilde{H}_{\text{pair}} $ is equivalent to the operator $4MJ_{x} + F$, where $F$ is a self-adjoint, bounded operator which is a nonlinear function of the $\mathfrak{su}(2)$ generators $J_{x}$, $J_{y}$, and $J_{z}$. If $F$ is neglected, one obtains an unparametrized approximate ground state $\ket{\tilde{\Phi}}$ of $\tilde{H}_{\text{pair}}$ from the $SU(2)$ coherent state ground state of $J_{x}$. Because the state $\ket{\tilde{\Phi}}$ allows to gain insight into several properties of the PC regime by analytical methods, we now show how to obtain it.

The unparametrized state $\ket{\tilde{\Phi}}$ is derived by considering an eigenvector of $J_{x}$ with eigenvalue $N/2$, i.e., proportional to
\begin{equation}
\bigl(b_{0}^{\dagger}+b_{1}^{\dagger}\bigr)^{N}\ket{0,0}.\nonumber
\end{equation}
in some spin-$N/2$ representation of $\mathfrak{su}(2)$.
By noting that, e.g.,
\begin{equation}
b_{1}^{\dagger}\ket{s,M-s}=\sqrt{M-s+1}\ket{s,M-s+1} \nonumber
\end{equation}
and
\begin{eqnarray}
\left( {n_{\tilde{1}} + n_{\tilde{3}} \over 2}  \right)^{-{1\over 2}}&{}&c_{\tilde{1}}^{\dagger}c_{\tilde{3}}^{\dagger}\ket{s,M-s,s,M-s} \nonumber \\ = &{}&\sqrt{M-s+1}\ket{s,M-s+1,s,M-s+1}, \nonumber
\end{eqnarray}
we see that $b_{1}^{\dagger}$ and $( ({n_{\tilde{1}} + n_{\tilde{3}}) / 2} )^{-{1\over 2}}c_{\tilde{1}}^{\dagger}c_{\tilde{3}}^{\dagger}$ are equivalent as linear operators and we can consider the state
\begin{eqnarray}\ket{\tilde{\Phi}} =  {1\over 2^{M/2}\sqrt{M!}}&{}&\left[  \left( {n_{\tilde{0}} + n_{\tilde{2}}\over 2}  \right)^{-{1\over 2}}c_{\tilde{0}}^{\dagger}c_{\tilde{2}}^{\dagger} \right. \nonumber \\  &{}&\left. +\left( { n_{\tilde{1}} + n_{\tilde{3}} \over 2}\right)^{-{1\over 2}}c_{\tilde{1}}^{\dagger}c_{\tilde{3}}^{\dagger}   \right]^{M}\ket{0,0,0,0}
\label{eqn:uuu}
\end{eqnarray}
as an unparametrized approximate ground state of $\tilde{H}_{\text{pair}}$. Transforming back to the original $\lbrace a_{j} \rbrace_{j=0,\ldots ,3}$ modes by $\mathcal{V}^{\dagger}$ produces a normalized model ground state $\ket{\Phi}$ for $H_{\text{pair}}$ given by
\begin{align}
\ket{\Phi}=&{1\over 2^{M}\sqrt{M!}}\left[ \left( {1\over \sqrt{n_0 + n_2}}\, \right)\left( a_{0}^{\dagger 2} + a_{2}^{\dagger 2} \right) \nonumber \right. \\
&+\left.\left( {1\over \sqrt{n_1+n_3}} \right) \left( a_{1}^{\dagger 2} + a_{3}^{\dagger 2} \right) \right]^{M}\ket{0,0,0,0} . \label{eqn:modelvar}
\end{align}
We refer to $\ket{\Phi}$ as a model ground state for $H_{\text{pair}}$.
Athough the model ground state in Eq.(\ref{eqn:modelvar}) can be considered as a condensate of $M$ pairs of particles, it is in stark contrast with pure Bose-Einstein condensate of $N=2M$ particles. In particular, a pure Bose-Einstein condensate has the form of $\ket{\omega}^{\otimes N}$  where $\ket{\omega}$ is a single particle state, and thus is a separable state under a particle partition; in contrast Eq.(\ref{eqn:modelvar}) is non-separable under a particle partitioning. The high-fidelity nature of $\ket{\Phi}$ is demonstrated in Fig.\ref{fig:fig1} where we find the fidelity $| \langle \Phi | \Psi_0\rangle|$ to the numerical ground state of $H_{\text{pair}}$ to be better than $0.99$ for up to $N=\mathcal{O}(10^3)$. Furthermore, the inset of Fig.\ref{fig:fig1} shows that the fidelity of $\ket{\Phi}$ can be reliably extrapolated to large $N$; for $N=\infty$ we find the fidelity $F_\infty \simeq 0.99256$. We demonstrate below that the entanglement properties of $\ket{\Phi}$ approximate those of the true ground state of $H_{\text{pair}}$ and use (\ref{eqn:modelvar}) as a model state to semi-quantitatively analyze accessible entanglement in the PC regime without the need for performing a variational optimization.

\begin{figure}
\includegraphics{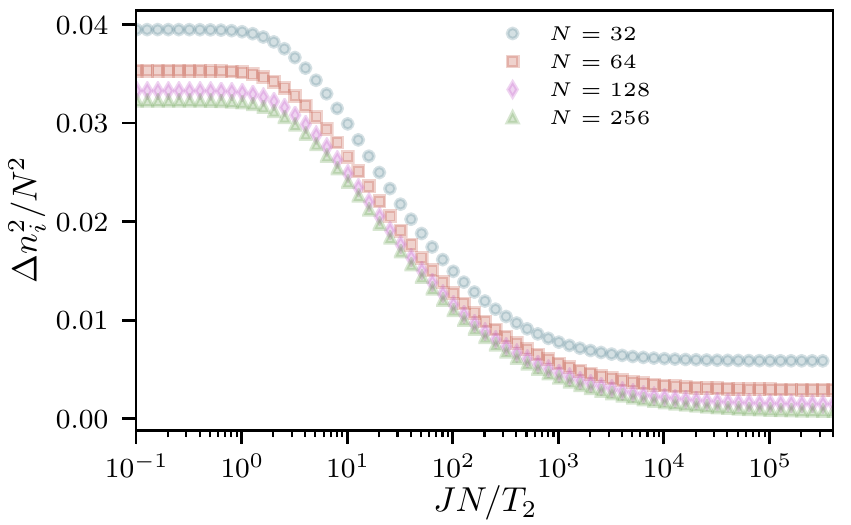}
\caption{\label{fig:fig2} On-site particle number variance $\langle \Delta n_i^{2}\rangle$ in the ground state of Eq.(\ref{eqn:ham}) with $U=0$ as a function of single particle tunneling strength, $J$ for different numbers of particles $N$.  In the pair-correlated regime, $\langle \Delta n_i^{2} \rangle \sim N^2$, whereas in the superfluid regime $\langle \Delta n_i^{2} \rangle \sim N$. This suggests that PC regime is stable up to $T_{2} \sim NJ$}
\end{figure}

In order to distinguish the three dynamical regimes of Eq.(\ref{eqn:ham}), the one-site occupation number variance $\langle \left( \Delta n_{j}\right)^{2} \rangle$ can be used as a local order parameter. Deep inside the Mott insulating regime, the variance vanishes, whereas in the superfluid regime the variance scales linearly with $N$, approaching $3N/16$ in the $U=T_2=0$ limit. In contrast, in Fig.\ref{fig:fig2}, we see that $\langle \left( \Delta n_{j}\right)^{2} \rangle$ scales as $\mathcal{O}(N^{2})$ in the PC regime. We can understand this by considering the model ground state $\ket{\Phi}$ for which
\begin{equation}
\langle \Phi | \left( \Delta n_{j}\right)^{2}| \Phi\rangle=\frac{1}{16}\left ({\frac{N^{2}}{2}}-N\right). \nonumber
\end{equation} Because $H_{\text{pair}}$ is quartic in the bosonic annihilation and creation operators, one expects that the ground state energy $\vert E_{0}(N) \vert$ scales as $\mathcal{O}(N^{2})$ and that a transition between superfluid and PC regime occurs in the regime $T_{2}/NJ \in \mathcal{O}(1)$. The numerical values of the local particle number variance in Fig.\ref{fig:fig2} are in agreement with a transition in this regime.  To verify the ground state energy scaling of $H_{\text{pair}}$, we show in Appendix \ref{sec:gse} that in the $N\rightarrow \infty$ limit, ${E_{0}(N)/ N^{2}} \le -1/2$.

\section{Accessible entanglement generation} 
\begin{figure}
\includegraphics{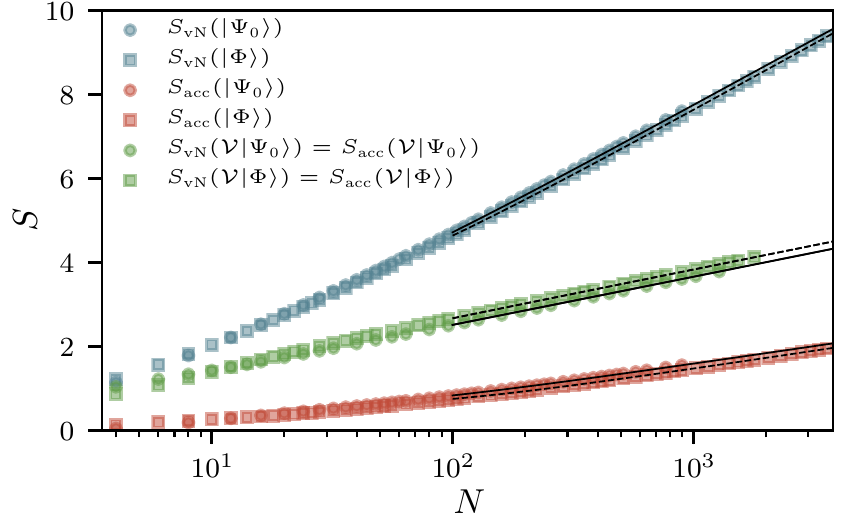}
\caption{\label{fig:fig3} Scaling of von-Neumann entanglement entropy $\Svn$ and accessible entanglement entropy $\Sacc$ with number of particles in the system $N$ for the ground state $\ket{\Psi_0}$ of Eq.(\ref{eqn:ham}) in the pure pair-tunneling limit ($U=J=0$) as well as the model ground state $\ket{\Phi}$ (Eq. \ref{eqn:modelvar} ), as well as both states after application of the beam-splitter operator $\mathcal{V}$. Note that $\Svn = \Sacc$ for both states $\mathcal{V}\ket{\Psi_0}$ and $\mathcal{V}\ket{\Phi}$. The solid (dashed) lines represent fits to $S = a \ln N + b + c/N$ for $\ket{\Psi_0}$ ($\ket{\Phi}$) .}
\end{figure}

We now consider the entanglement properties of the ground state $\ket{\Psi_{0}}$ of Eq.(\ref{eqn:ham}) in the pure pair-tunneling limit. We bipartition the system into neighboring pairs of modes ($\lbrace 0,1 \rbrace \sqcup \lbrace 2,3 \rbrace$) and quantify the entanglement between these pairs of modes with the entanglement entropy $\Svn$ and accessible entanglement entropy $\Sacc$. Fig.~\ref{fig:fig3} shows the scaling of $\Svn$ and $\Sacc$ with $N$ for both $\ket{\Psi_{0}}$ and the model ground state $\ket{\Phi}$ in Eq.(\ref{eqn:modelvar}). For both the model ground state and the exact ground state, we find that both $\Svn$ and $\Sacc$ scale as $\ln N$; the lines in Fig.~\ref{fig:fig3} represent three parameter fits to $S = a\ln N + b + c/N$; for $\ket{\Psi_0}$ we find $a_{\text{vN}} \simeq 1.36$ and $a_{\text{acc}} \simeq 0.37$, and for $\ket{\Phi}$, $a_{\text{vN}} \simeq 1.37$ and $a_{\text{acc}} \simeq 0.38$. We see that $\ket{\Phi}$ quantitatively captures the entanglement in the  pure pair-tunneling limit. Although the majority of the entanglement is due to fluctuations between subsystems and thus inaccessible via local operations, a finite fraction of entanglement remains accessible in the large $N$ limit.

To increase the accessible entanglement, consider again the unitary operator $\mathcal{V}$ from Sec. \ref{sec:pcssec}. With respect to the mode bipartition $\lbrace 0,1 \rbrace \sqcup \lbrace 2,3 \rbrace$, $\mathcal{V}$ is a non-local  operation. Such a non-local mode transformation can be implemented experimentally via, e.g., a matter wave beamsplitter \cite{schmied}. If the four modes are arranged in a tetrahedral configuration in a 3-D optical lattice \cite{greiner3d} with tunable tunnel couplings so as to generate the ring topology, the beam splitter $\mathcal{V}$ can be implemented via dynamical potential splitting \cite{schmied} or a coherent Y-junction splitter \cite{boshier}, either method could implement a matter-wave beamsplitter between any chosen pair of modes due to the tetrahedral symmetry.

We proceed to consider the entanglement of $\mathcal{V}\ket{\Psi_{0}}$ under the bipartition $\lbrace \tilde{0},\tilde{1} \rbrace \sqcup \lbrace \tilde{2},\tilde{3} \rbrace$ of the new modes. After the implementation of $\mathcal{V}$, it is observed that $\Sacc=\Svn$ and correspondingly all entanglement is operationally accessible by local operations. Physically, this entanglement conversion is due to the absence of particle number fluctuations between the $\lbrace \tilde{0},\tilde{1} \rbrace $ and $\lbrace \tilde{2},\tilde{3} \rbrace$ modes; in particular, $\Sfluc = 0$ for both $\mathcal{V} \ket{\Phi}$ and $\mathcal{V}\ket{\Psi_{0}}$ since $\mathcal{V}$ maps $\ket{\Phi}$ and $\ket{\Psi_{0}}$ to states with exactly $N/2$ particles in each subset of modes. Additionally, the coefficient of the $\ln N$ scaling of $\Sacc$ exhibits a switch-like increase (Fig.\ref{fig:fig3}) upon the operation of $\mathcal{V}$ on the ground state $\ket{\Psi_{0}}$. In particular, we find that $\Sacc(\mathcal{V}\ket{\Psi_{0}})$ is best fit with coefficient $a_{\text{acc}} \simeq 0.50$ of the leading logarithmic scaling. In Appendix \ref{sec:app3}, we derive an upper bound on $\Sacc( \mathcal{V}\ket{\psi(c)})$, the accessible entanglement of the variational ground state of $\tilde{H}_{\text{pair}}$, from which the observed logarithmic scaling of $\Sacc(\mathcal{V}\ket{\Psi_{0}})$ in Fig.\ref{fig:fig3} can also be bounded due to the exactness of the variational subspace.

As in the case of $\ket{\Phi}$ and $\ket{\Psi_{0}}$, it is clear from Fig.\ref{fig:fig3} that the transformed model ground state $\ket{\tilde{\Phi}}=\mathcal{V}\ket{\Phi}$ faithfully approximates the entanglement scaling of the exact ground state $\mathcal{V}\ket{\Psi_{0}}$ of $\tilde{H}_{\text{pair}}$. The $\mathcal{O}(\ln N)$ scaling of $\Sacc(\mathcal{V}\ket{\Psi_{0}})$, the accessible entanglement of the transformed ground state deep in the PC regime, can be understood by an analytical calculation of $\Sacc(\mathcal{V}\ket{\Phi})$. Consider the expression for $\mathcal{V}\ket{\Phi}$ in the Fock basis (see Eq.(\ref{eqn:uuu})):
\begin{equation}
\mathcal{V}\ket{\Phi} = {1\over 2^{M/2}}\sum_{j=0}^{M}\sqrt{ {M\choose j}} \ket{j,M-j,j,M-j}. \nonumber
\end{equation}
Since for $\mathcal{V}\ket{\Phi}$ the number of particles in each subset of modes under this bipartition are the same, $\Sfluc=0$ and $\Svn = \Sacc$. From the above expression, it is clear that $\Svn(\mathcal{V}\ket{\Phi})$ is equal to the Shannon entropy of a random variable that obeys the $B(M=N/2,p=1/2)$ binomial distribution:
\begin{equation}
\Svn(\mathcal{V}\ket{\Phi}) =-\sum_{j=0}^{M}{ {M\choose j}\over 2^{M}} \ln { {M\choose j}\over 2^{M}}. \nonumber
\end{equation}

It then follows from the central limit theorem that as $N\rightarrow \infty$, $\Sacc(\mathcal{V}\ket{\Phi}) ={1\over 2}\ln N + \mathcal{O}(1)$, which is consistent with the scaling of $\Sacc(\mathcal{V}\ket{\Psi_{0}})$ stated above. From this analysis, it follows that the unitary beamsplitter operation $\mathcal{V}$  converts $\ket{\Psi_{0}}$ into a state with completely accessible entanglement, and furthermore results in an increase of the $\ln N$ scaling of the accessible entanglement $\Sacc$ deep in the PC regime. This increase in the coefficient of the $\ln N$ scaling of $\Sacc$ brought about by $\mathcal{V}$ occurs for both the exact ground state and model ground state.

\section{Conclusion} Through analyses of high-fidelity variational states and parameter-free approximate ground states, we have shown that accessible entanglement can be generated via pair-correlated tunneling in an extended Bose-Hubbard model. We have thus demonstrated that pair-correlated tunneling can drive many-boson systems into states with entanglement that can be locally accessed for quantum information processing protocols. Additionally, by implementing a matter wave beamsplitter, the ground state of the PC regime is transformed into a state with fully accessible entanglement, i.e., a state in which all entanglement has been concentrated into a single particle number sector. The accessible entanglement switching behavior in the four-mode system considered in this work, which is the minimal mode number for which $\Sacc > 0$, complements recent results on using three- and four-mode bosonic models to analyze matter-wave entanglement dynamics \cite{PhysRevLett.106.120405,fourwavemix,fourwavemix2,fourwavemix3,fourwavemix4}. We expect that the existence of a low-complexity variational ground state subspace and high-fidelity model wavefunctions for the PC regime will stimulate further analyses of the quantum information processing capabilities of the PC regime, including the interplay with other quantum phenomena such as superfluidity \cite{tjvyk}.

\acknowledgments
The numerical diagonalization code used for this work was based on a code developed by R. Melko and D. Iouchtchenko~\cite{melkoherdman}. TJV is supported by the Korea Research Fellowship Program through the National Research Foundation of Korea (NRF) funded by the Ministry of Science and ICT (2016H1D3A1908876) and by the Basic Science Research Program through the National Research Foundation of Korea (NRF) funded by the Ministry of Education (2015R1D1A1A09056745).

\appendix

\section{\label{sec:app1}Normalization and inner products of the variational basis states}

Here we will consider the normalization of the basis states $\ket{\varphi_{\ell}}$ of the variational ground-state subspace of $H_{\text{pair}}$.

\textbf{Proposition.} \textit{If $k_{\ell} \in \lbrace 0,\pi \rbrace$, then \begin{equation}\mathcal{N}_{\ell}=2^{M+1}M!\sqrt{M+1}.
\nonumber\end{equation} If $k_{\ell} \notin \lbrace 0,\pi \rbrace$, then \begin{equation} \mathcal{N}_{\ell}=2^{M}M!\sqrt{2M+4}.\nonumber\end{equation}}

\textbf{Proof.}
Let \begin{align}
\ket{\beta_{\ell} }&=\left[ \vphantom{a\over b}\left( a_{0}^{\dagger 2}+e^{ik_{\ell}}a_{1}^{\dagger 2}+a_{2}^{\dagger 2}+e^{ik_{\ell}}a_{3}^{\dagger 2} \right)^{M} \right. \nonumber \\  &+ \left. \left( a_{0}^{\dagger 2}+e^{-ik_{\ell}}a_{1}^{\dagger 2}+a_{2}^{\dagger 2}+e^{-ik_{\ell}}a_{3}^{\dagger 2} \right)^{M}  \vphantom{a\over b} \right] \ket{0,0,0,0}  \nonumber
\label{eqn:betastate}
\end{align}
be an \textit{ unnormalized} superposition of paired states. Under the action of $\mathcal{V}$ in Eq.(6) of the main text, the state
\begin{equation}
\left( a_{0}^{\dagger 2}+e^{ik_{\ell}}a_{1}^{\dagger 2}+a_{2}^{\dagger 2}+e^{ik_{\ell}}a_{3}^{\dagger 2} \right)^{M} \ket{0,0,0,0} \nonumber
\end{equation}
 is transformed isometrically to
 \begin{equation}
 2^{M}\left( c_{\tilde{0}}^{\dagger}c_{\tilde{2}}^{\dagger} + e^{ik_{\ell}}c_{\tilde{1}}^{\dagger}c_{\tilde{3}}^{\dagger} \right)^{M}\ket{0,0,0,0}. \nonumber
 \end{equation}
Using the binomial theorem, one finds that the states
\begin{equation}
\ket{\xi_{\pm\ell}}={1\over M! \sqrt{M+1}}\left( c_{\tilde{0}}^{\dagger}c_{\tilde{2}}^{\dagger} + e^{\pm ik_{\ell}}c_{\tilde{1}}^{\dagger}c_{\tilde{3}}^{\dagger} \right)^{M}\ket{0,0,0,0} \nonumber
\end{equation}
are normalized, where $\ell \in \lbrace 0,1,\ldots , \lfloor M/2 \rfloor \rbrace$. If $k_{\ell}\in \lbrace 0 , \pi \rbrace$, then
\begin{equation}
\ket{\beta_{\ell} } = 2\left( a_{0}^{\dagger 2}+e^{ik_{\ell}}a_{1}^{\dagger 2}+a_{2}^{\dagger 2}+e^{ik_{\ell}}a_{3}^{\dagger 2} \right)^{M}\ket{0,0,0,0} \nonumber
\end{equation}
gets mapped to
\begin{equation}
2^{M+1}\left( c_{\tilde{0}}^{\dagger}c_{\tilde{2}}^{\dagger} + e^{ik_{\ell}}c_{\tilde{1}}^{\dagger}c_{\tilde{3}}^{\dagger} \right)^{M}\ket{0,0,0,0} = 2^{M+1}M!\sqrt{M+1}\ket{\xi_{\ell}}, \nonumber
\end{equation}
and, therefore,
\begin{equation}
\ket{\varphi_{\ell}}=\frac{1}{2^{M+1}M!\sqrt{M+1}}\ket{\beta_{\ell} } \nonumber
\end{equation}
is normalized.

To calculate $\mathcal{N}_{\ell}$ for $k_{\ell}\notin \lbrace 0, \pi \rbrace$, it is necessary to calculate the inner product $\langle \xi_{q} \vert \xi_{\ell} \rangle$ by using the binomial theorem.
For any $q \neq \ell$ the result is
\begin{eqnarray}
\langle \xi_{q} \vert \xi_{\ell} \rangle &=&{1\over M+1} \sum_{j=0}^{M}\sum_{s=0}^{M}\left[ \vphantom{{a\over b}} (e^{-ik_{q}})^{M-j}  (e^{ik_{\ell}})^{M-s} \right. \nonumber \\   &{}& \left. \langle j,M-j,j,M-j \vert s,M-s,s,M-s \rangle \vphantom{{a\over b}} \right] \nonumber \\
&=& {1\over M+1}\sum_{j=0}^{M}(e^{i(k_{\ell}-k_{q})})^{M-j} \nonumber \\
&=& {1\over M+1}.
\label{eqn:calcn}
\end{eqnarray}
where $\ket{n_{0},n_{1},n_{2},n_{3}}$ are Fock states in the $c_{\tilde{j}}^\dagger$ basis.

Now we normalize $\ket{\beta_{\ell}}$ for $k_{\ell} \notin \lbrace 0,\pi \rbrace$. Note that $\ket{\beta_{\ell}}$ is transformed isometrically to
\begin{equation}
2^{M}M!\sqrt{M+1}\biggl( \ket{\xi_{\ell}} + \ket{\xi_{-\ell}} \biggr) \nonumber
\end{equation}
under the rotation $\mathcal{V}$, and that
\begin{equation}
{\sqrt{M+1}\over\sqrt{2M+4}}\biggl( \ket{\xi_{\ell}} + \ket{\xi_{-\ell}} \biggr) \nonumber
\end{equation}
is a normalized state. Therefore,
\begin{equation}\ket{\varphi_{\ell}} = {1\over 2^{M}M!\sqrt{2M+4}}\ket{\beta_{\ell}} \nonumber
\end{equation} is normalized.\hspace{1cm}$\square$

The inner products of the normalized states $\ket{\varphi_{\ell}}$, or, equivalently, their Gram matrix, are the subject of the following Proposition.

\textbf{Proposition.} \textit{If $k_{\ell}\in \lbrace 0,\pi \rbrace$ and $k_{r} \in \lbrace 0,\pi \rbrace$ and $\ell \neq r$, then
\begin{equation}
\langle \varphi_{\ell} \vert \varphi_{r} \rangle = {1\over M+1}. \nonumber
\end{equation}
If $k_{\ell} \in \lbrace 0,\pi \rbrace$ and $k_{r} \notin \lbrace 0,\pi \rbrace$, then
\begin{equation}
\langle \varphi_{\ell} \vert \varphi_{r} \rangle = \sqrt{2\over (M+1)(M+2)}.
\label{eqn:middle}
\end{equation}
If $k_{r}\notin \lbrace 0,\pi \rbrace$ and $k_{\ell} \notin \lbrace 0,\pi \rbrace$ and $\ell \neq r$, then
\begin{equation}
\langle \varphi_{\ell} \vert \varphi_{r} \rangle = {2\over M+2}. \nonumber
\end{equation}}

\textbf{Proof.} We prove Eq.(\ref{eqn:middle}) explicitly and note that the other inner products are proved in the same way. If $k_{\ell} \in \lbrace 0,\pi \rbrace$, then
\begin{equation}
\ket{\varphi_{\ell}}={1\over 2^{M+1}M!\sqrt{M+1}}\ket{\psi_{\ell}}\nonumber
\end{equation}
gets mapped isometrically to $\ket{\xi_{\ell}}$ under the action of $\mathcal{V}$. If $k_{r} \notin \lbrace 0,\pi \rbrace$, then
\begin{equation}
\ket{\varphi_{r}}=  {1\over 2^{M}M!\sqrt{2M+4}}\ket{\psi_{r}}\nonumber
\end{equation}
gets mapped isometrically to
\begin{equation}
{\sqrt{M+1}\over\sqrt{2M+4}}\biggl( \ket{\xi_{r}} + \ket{\xi_{-r}} \biggr)\nonumber
\end{equation}
under the action of $\mathcal{V}$. Therefore, by Eq.(\ref{eqn:calcn}),
\begin{align}
\langle \varphi_{\ell} \vert \varphi_{r} \rangle &= {\sqrt{M+1}\over\sqrt{2M+4}}\biggl( \langle \xi_{\ell} \vert \xi_{r} \rangle + \langle \xi_{\ell} \vert \xi_{-r} \rangle \biggr)  \nonumber\\
&= \sqrt{2\over (M+1)(M+2)}. \nonumber
\end{align}
\hspace{1cm} $\square$

\section{\label{sec:gse}Ground state energy of PC regime}

To obtain a lower bound for the particle number-dependent ground state energy $E_{0}(N)$ in the PC regime, one can consider the expectation of the operator given in Eq. \ref{eqn:Hpairspin} in the state $\ket{J_{x} = -M/2}$, which defines the eigenvector of $J_{x}$ with eigenvalue $-M/2$. One finds that
\begin{equation}
\langle \tilde{H}_{\text{pair}} \rangle_{\ket{J_{x} = -M/2}} = -2M^{2} +\mathcal{O}(M).\nonumber
\end{equation}
Therefore,
\begin{equation}
E_{0}(N)\le \langle  \tilde{H}_{\text{pair}} \rangle_{\ket{J_{x} = -M/2}}= -{1\over 2}N^{2} + p(N),\nonumber
\end{equation}
where $p(N)$ is a polynomial which is linear in $N$. It follows that
\begin{equation}
\lim_{N\rightarrow \infty} {E_{0}(N) \over N^{2}} \le \lim_{N\rightarrow \infty} { \langle \tilde{H}_{\text{pair}} \rangle_{\ket{J_{x} = -M/2}} \over N^{2}} = -{1\over 2}. \nonumber
\end{equation}
\vspace{.5cm}
\section{\label{sec:app3}Upper bound on $\Sacc( \mathcal{V}\ket{\psi(c)})$}

The analytical expression for $\Sacc(\ket{\psi(c)})$ and $\Sacc(\ket{\Phi})$, i.e., the accessible of the variational ground state and model ground state, respectively, in the PC regime, with respect to the $\lbrace 0,1 \rbrace \cup \lbrace 2,3 \rbrace$ bipartition is cumbersome and will not be shown here. However, an upper bound for the accessible entanglement of the variational ground state $\mathcal{V}\ket{\psi(c)}$ can be computed by utilizing the inequality for a state  $\ket{\psi}$
\begin{equation}
\Sacc\bigl(\ket{\psi}\bigr) \le -\sum_{\substack{n=0\\ p_{n}\neq 0}}^{N}p_{n}\ln \text{Tr} \left( {1  \over p_{n}}\text{Tr}_{B}P_{n}\ket{\psi}\bra{\psi}P_{n} \right)^{2} \label{eqn:Svnbound}
\end{equation}
where the right hand side involves an average of 2nd R\'{e}nyi entropies in for $\ket{\psi}$ projected onto states with definite number sector in each subsystem.

For the transformed variational ground state $\mathcal{V}\ket{\psi(c)}$, the right-hand side of Eq. \ref{eqn:Svnbound} can be easily calculated for  either of the partitions $\lbrace \tilde{0},\tilde{1} \rbrace \cup \lbrace \tilde{2},\tilde{3} \rbrace$ or $\lbrace \tilde{0},\tilde{2} \rbrace \cup \lbrace \tilde{1},\tilde{3} \rbrace$. In particular,  for $\lbrace \tilde{0},\tilde{1} \rbrace \cup \lbrace \tilde{2},\tilde{3} \rbrace$ , making use of the relation
\begin{widetext}
\begin{align}
\mathcal{V}\ket{\psi(c)} =& c_{0}\ket{\xi_{0}}+c_{M/2}\ket{\xi_{M/2}} + \sum_{\ell =1}^{\lfloor M/2 \rfloor -1}c_{\ell}\sqrt{{M+1 \over 2M+4}}\left( \ket{\xi_{\ell}}+\ket{\xi_{-\ell}} \right) \nonumber
\end{align}
and the fact that
\begin{equation}
\text{tr}_{\tilde{2},\tilde{3} }P_{n}\ket{\xi_{\ell}}\bra{\xi_{\ell'}}P_{n} = {\delta_{n,M} \over M+1} \sum_{j=0}^{M}e^{i(k_{\ell'}-k_{\ell})j}\ket{j,M-j}\bra{j,M-j} , \nonumber
\end{equation} gives the result
\begin{eqnarray}
\Sacc\bigl(\mathcal{V}\ket{\psi(c)}\bigr) &\le& -\ln\sum_{j=0}^{M}\left( \vphantom{\sum_{\ell,\ell' =1}^{\lfloor M/2 \rfloor -1}} {\left(c_{0}+(-1)^{j}c_{M/2}\right)^{2}\over M+1} + \sum_{\ell=1}^{\lfloor M/2 \rfloor -1} {4c_{\ell}\cos(k_{\ell}j)\left( c_{0}+(-1)^{j}c_{M/2} \right)\over \sqrt{(2M+4)(M+1)}} \right. \nonumber \\ &+& \left. \sum_{\ell,\ell' =1}^{\lfloor M/2 \rfloor -1}{ c_{\ell}c_{\ell'}2\cos(k_{\ell}j)\cos(k_{\ell'}j) \over M+2} \right)^{2}\label{eqn:s2op}
\end{eqnarray}
\end{widetext}
where $c_{M/2}=0$ if $N \neq 0 \text{ mod } 4$. The same method can be used to show that $\Sacc(\mathcal{V}\ket{\psi(c)})=0$ for the $\lbrace \tilde{0},\tilde{2} \rbrace \cup \lbrace \tilde{1},\tilde{3} \rbrace$ partition. Futhermore, because $\mathcal{V}$ is a tensor product of beamsplitters taking $\lbrace 0,2 \rbrace \rightarrow \lbrace \tilde{0},\tilde{2} \rbrace$ and $\lbrace 1,3 \rbrace \rightarrow \lbrace \tilde{1},\tilde{3} \rbrace$,  $\Sacc(\ket{\psi(c)})$ also vanishes for the $\lbrace 0,2 \rbrace \cup \lbrace 1,3 \rbrace$ bipartition. Thus we conclude that the accessible entanglement of the ground state of $H_{\text{pair}}$ in the pure pair tunneling limit vanishes when the $\lbrace 0,2 \rbrace \cup \lbrace 1,3 \rbrace$ bipartition is considered.

\bibliography{pairtunnbib.bib}

\end{document}